\documentclass[prd,aps,twocolumn,floats,floatfix,nofootinbib]{revtex4-1}
\usepackage{graphicx}
\usepackage{amsmath,amssymb}
\usepackage{physics}
\usepackage{subfigure}
\usepackage{appendix}
\usepackage[table]{xcolor}
\usepackage{hyperref}
\usepackage{booktabs} 
\usepackage{longtable} 
\usepackage{enumerate}
\begin{document}

\title{Modeling the postmerger gravitational wave signal and extracting binary properties 
from future binary neutron star detections}

\author{
Ka Wa Tsang$^{1,2}$, 
Tim Dietrich$^{1}$, 
Chris Van Den Broeck$^{1,2}$
}

\affiliation{$^1$Nikhef -- National Institute for Subatomic Physics, 105 Science Park, 
1098 XG Amsterdam, The Netherlands \\
$^2$Van Swinderen Institute for Particle Physics and Gravity, University of Groningen, \\
Nijenborgh 4, 9747 AG Groningen, The Netherlands}

\date{\today}


\begin{abstract}
Gravitational wave astronomy has established its role in 
measuring the equation of state governing cold supranuclear matter. 
To date and in the near future, gravitational wave measurements from 
neutron star binaries are likely to be restricted to the inspiral. 
However, future upgrades and the next generation of gravitational wave 
detectors will enable us to detect the gravitational wave signatures
emitted after the merger of two stars, at times when 
densities beyond those in single neutron stars are reached. 
Therefore, the postmerger gravitational wave signal enables studies of 
supranuclear matter at its extreme limit. 
To support this line of research, we present new and updated phenomenological
relations between the binary properties and characteristic features of the postmerger evolution. 
Most notably, we derive an updated relation connecting the mass-weighted tidal deformability and 
the maximum neutron star mass to the dominant emission frequency of the 
postmerger spectrum. 
With the help of a configuration-independent Bayesian analysis 
using simplified Lorentzian model functions, we find 
that the main emission frequency of the postmerger remnant, 
for signal-to-noise ratios of $8$ and above, can be extracted 
within a 1-sigma uncertainty of about 100 Hz for 
Advanced LIGO and Advanced Virgo's design sensitivities. 
In some cases, even a postmerger signal-to-noise ratio of $4$ can be sufficient 
to determine the main emission frequency. 
This will enable us to measure binary and equation-of-state properties from the postmerger, 
to perform a consistency check between different parts of the binary neutron star 
coalescence, and to put our physical interpretation of neutron star mergers to the test. 
\end{abstract}

\maketitle

\section{Introduction}
\label{sec:introduction}

The extreme densities and conditions inside neutron stars (NSs) 
cannot be reached in existing experiments. This makes 
NSs a unique laboratory to study the equation 
of state (EOS) governing cold-supranuclear dense material. 
Following the first detection of a gravitational wave (GW) signal 
originating from the coalescence of a binary neutron star (BNS) 
system, GW170817, by the Advanced LIGO~\cite{TheLIGOScientific:2014jea}
and Advanced Virgo detectors~\cite{TheVirgo:2014hva},
it became possible to constrain the NS EOS 
by analyzing the measured GWs~\cite{TheLIGOScientific:2017qsa,Abbott:2018wiz,
Abbott:2018exr,LIGOScientific:2018mvr,De:2018uhw}.
Because of the increasing sensitivity of GW interferometers,
multiple detections of merging BNSs are expected
in the near future~\cite{Aasi:2013wya}. This will make 
GW astronomy an inevitable tool within nuclear physics 
community. 

In general, there are two ways to extract information 
about the EOS governing the NS's interior from a GW 
detection. 
The first method relies on the modeling of the BNS 
inspiral~\cite{Hinderer:2009ca,Damour:2012yf,DelPozzo:2013ala,
Lackey:2014fwa,Agathos:2015uaa} and on waveform approximants that 
include tidal effects, represent accurately the system's properties, 
and are of sufficiently low computational cost that they can be used  
in parameter estimation pipelines, 
e.g.,~\cite{Dietrich:2018uni,Kawaguchi:2018gvj,Damour:2012yf}. 
The zero-temperature EOS is then constrained by measuring 
a mass-weighted combination of the quadrupolar tidal deformability 
$\tilde{\Lambda}$ or similar parameters that characterize tidal 
interactions, e.g.,~\cite{Hinderer:2007mb,Damour:2009wj}. 

The second method relies on an accurate modeling of 
the postmerger GW spectrum, e.g.,~\cite{Bauswein:2011tp,Clark:2014wua,Takami:2014zpa,
Rezzolla:2016nxn,Bernuzzi:2015rla}, and can deliver an independent estimate of the EOS
at densities exceeding the ones present in single NSs~\cite{Radice:2016rys}. 
Therefore, the postmerger modeling also allows to investigate 
interesting phenomena such as phase transitions happening inside the merger 
remnant at very high densities~\cite{Bauswein:2018bma,Most:2018eaw}. 
It is expected that all BNS merger remnants which do not undergo prompt 
collapse, e.g.,~\cite{Bauswein:2013jpa,Koppel:2019pys}, 
will radiate a significant amount of energy in the form of GWs~\cite{Bernuzzi:2015opx,Zappa:2017xba}. 
This radiation has a characteristic GW spectrum 
composed of a few peaks at frequencies $f_\text{GW}\sim2$-$4$~kHz. 
The main peak frequencies of the postmerger spectrum correlate 
to properties of a zero-temperature spherical equilibrium 
star as outlined in previous works, e.g.,
\cite{Bauswein:2011tp,Bauswein:2012ya,Hotokezaka:2013iia,
Bauswein:2014qla,Takami:2014zpa,Takami:2014tva,Clark:2014wua,
Bernuzzi:2015rla,Bauswein:2015yca,Rezzolla:2016nxn,Bose:2017jvk,
Chatziioannou:2017ixj,Easter:2018pqy,Torres-Rivas:2018svp}. 

To date, the advanced GW detectors have only been able to 
observe the inspiral of the two NSs~\cite{Abbott:2017dke,Abbott:2018hgk} and  
no postmerger signal has been observed. This non-observation is caused by the 
higher emission frequency at which current GW detectors are less sensitive.
But, the increasing sensitivity of the 2nd generation of GW detectors (Advanced LIGO 
and Advanced Virgo) will not only  
increase the detection rate of BNS inspiral signals, 
there will also be the chance of observing the postmerger signal for a few `loud' events.
Ref.~\cite{Dudi:2018jzn} finds that for sources similar to GW170817 but observed 
with Advanced LIGO and Advanced Virgo's design sensitivities, 
the postmerger part of the BNS coalescence might have an SNR of $\sim 2$-$3$. 
The planned third generation of GW interferometers, 
e.g., the Einstein Telescope~\cite{Hild:2008ng,Punturo:2010zz,Ballmer:2015mvn} 
or the Cosmic Explorer~\cite{Evans:2016mbw}, have the capability to detect 
the postmerger signal of upcoming BNS mergers with SNRs up to $\sim 10$.

Unfortunately, the postmerger spectrum is influenced in a complicated 
way by thermal effects, magnetohydrodynamical instabilities, 
neutrino emissions, phase transitions, and dissipative processes, 
e.g.,~\cite{Siegel:2013nrw,Alford:2017rxf,Radice:2017zta,
Shibata:2017xht,Bauswein:2018bma,Most:2018eaw,DePietri:2018tpx}. 
Currently, any postmerger study relies heavily on expensive numerical 
relativity (NR) simulations and there is to date no possibility to perform simulations 
incorporating all necessary microphysical processes. 
Therefore, our current theoretical understanding of this part of the BNS 
coalescence is overall limited. 
In addition, there has been no NR simulation yet which 
has been able to show convergence of the GW phase in the postmerger. 
While this observation can be generally explained by the presence of shocks or 
discontinuities formed during the collision of 
the two stars, it also increases our uncertainty 
on any quantitative result. 

Nevertheless, the community tried to construct postmerger approximants focusing on 
characteristic (robust) features present in NR simulations.
The discovery of quasi-universal relations is a building block for 
most descriptions of the postmerger GW spectrum.
Clark \emph{et al.}~\cite{Clark:2015zxa} showed that a principle component analysis 
can be used to reduce the dimensionality of the spectrum for equal mass binaries once
the different spectra are normalized and aligned such that the main emission 
frequencies coincide. 
Effort has also been put to model the plus polarization in time domain using a
superposition of damped sinusoids incorporating 
quasi-universal relations~\cite{Bose:2017jvk}.
Relying on a very accurate $f_2$ estimate for 
an accurate rescaling of the waveforms 
Easter \emph{et al.}~\cite{Easter:2018pqy} created a 
hierarchical model to estimate the postmerger spectra.

Here, we follow a similar path and try to describe the GW spectrum with a set of 
a three- and a six-parameter model function with a Lorentzian-like shape. 
Comparing our ansatz with a set of $54$ NR simulations,
we find average mismatches of $0.18$ 
for the three-parameter and $0.15$ 
for the six-parameter model; cf.~Tab.~\ref{tab:NR_data}.
Our approximants do not incorporate directly quasi-universal relations, but are constructed 
to describe generic postmerger waveforms. 
Thus, our analysis is flexible and allows to describe almost arbitrary configurations. 
Employing our model in standard parameter estimation pipelines~\cite{lalsuite,Veitch:2014wba}
of the LIGO and Virgo Collaborations, we find that we can extract the dominant emission 
frequency in the postmerger for a number of tests. 
To our knowledge, this is the first time a 
model-based (but configuration independent) method is employed within a 
Bayesian analysis of the postmerger signal. 

Once the individual parameters describing the postmerger spectra are extracted, 
we use fits for the peak frequency to connect the measured signal to the properties 
of the supranuclear EOS and the merging binary. 
This way, one can combine measurements from the inspiral and postmerger phase 
to provide a consistency test for our supranuclear matter description. 

Although not used here, we want to mention an alternative approach, 
which employs the morphology-independent
burst search algorithm called \texttt{BayesWave}~\cite{Cornish:2014kda,Becsy:2016ofp}.
Ref.~\cite{Chatziioannou:2017ixj} showed that this approach is capable of reconstructing 
the postmerger signal and allows to extract properties from the measured GW signal.
Even for a measured postmerger SNR of $\sim 5$, the main emission frequency of the remnant 
could be determined within a few dozens of Hz.
Compared to \texttt{BayesWave}, our simple model functions might have the advantage that 
without any modifications of the current code for statistical inferences,
in particular the \texttt{LALInference} module~\cite{Veitch:2014wba}
available in the LSC Algorithm Library (LAL) Suite,
they can be added to existing frequency domain inspiral-merger 
waveforms describing the first part of the BNS coalescence, 
e.g.,~\cite{Damour:2012yf,Dietrich:2018uni,Kawaguchi:2018gvj,Messina:2019uby,
Schmidt:2019wrl,Dietrich:2019kaq}, 
to construct a full inspiral-merger-postmerger (IMP) waveform directly employable for GW analysis.
Such an IMP study can also be carried out 
within the \texttt{BayesWave} approach, but seems technically harder since one has to combine
model-based and non-model-based algorithms. \\


Our paper is structured as follows.
In Sec.~\ref{sec:description}, we discuss the general time domain and frequency domain morphologies
of the postmerger signal as obtained from NR simulations.
Based on this discussion, we derive new quasi-universal relations
for the time between the merger and the first time domain amplitude minimum, and the first time domain amplitude maximum. 
We also extend existing quasi-universal relation for the main emission 
frequency $f_2$ of the GW postmerger spectrum and its amplitude in the frequency domain. 
In Sec.~\ref{sec:model_function} we discuss two different Lorentzian model functions 
and their performance to model NR simulations. 
In Sec.~\ref{sec:parameter_estimation} a full Bayesian analysis of 
a set of NR model waveforms is performed. 
In Sec.~\ref{sec:results} we show how our analysis can be used to constrain the EOS 
and how to test consistency between the inspiral and postmerger. 
We conclude in Sec.~\ref{sec:conclusions}.
We list in the appendix the NR data 
employed for the construction of the 
quasi-universal relations presented in the main text. 

Unless otherwise stated, 
this paper uses geometric units by setting $G=c=1$.
Throughout the work, we employ the NR simulations published
in the Computational Relativity (CoRe) database~\cite{Dietrich:2018phi}.
In addition, where explicitly mentioned,
we increase our dataset by adding results published in~\cite{Takami:2014tva, Hotokezaka:2013iia}.
We refer the reader to Tab.~\ref{tab:NR_data} for further details about the individual data. 

\section{The postmerger morphology}
\label{sec:description}

\subsection{Time Domain}

\begin{figure}[t]
 \centering
 \includegraphics[width=\columnwidth]{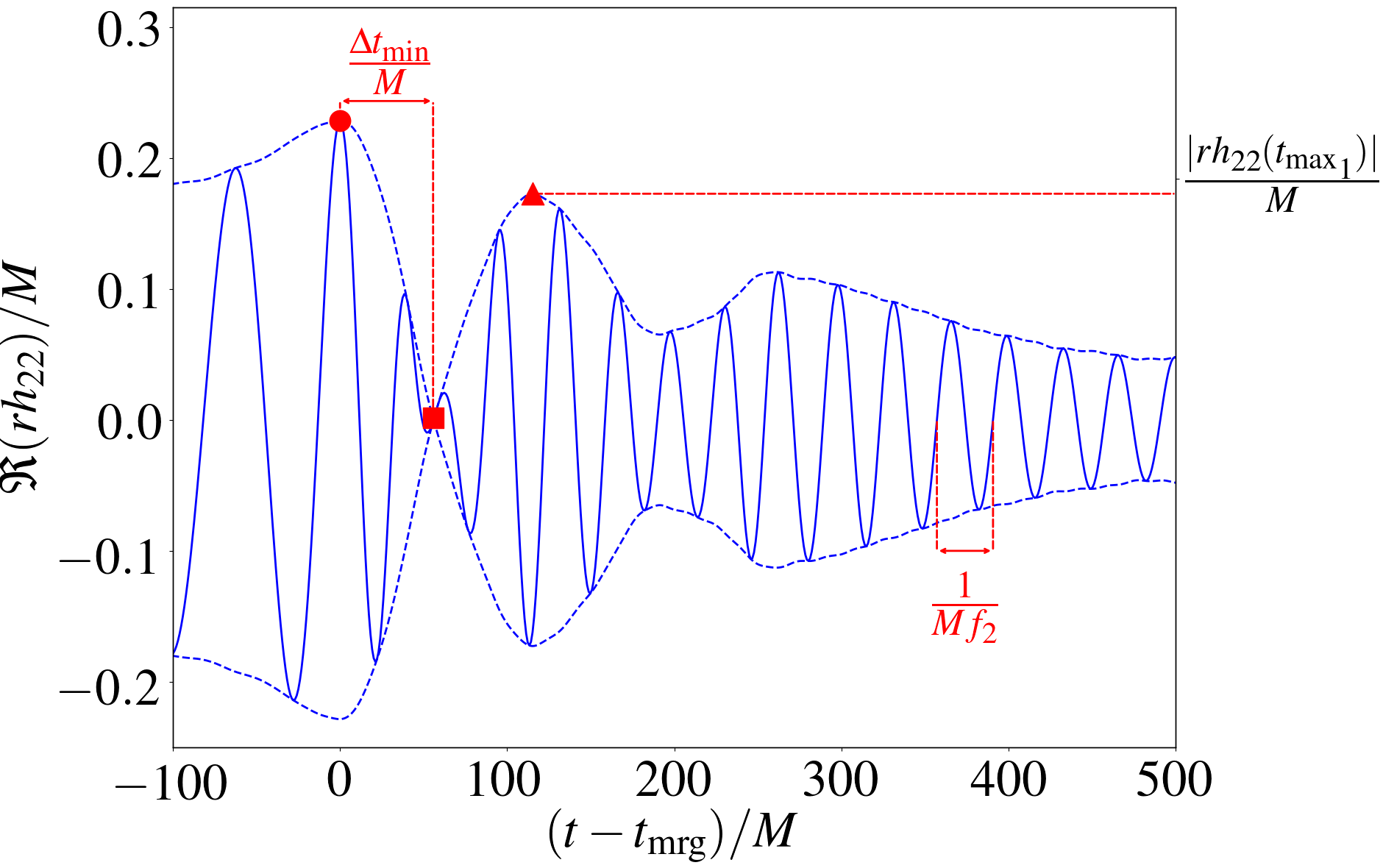}
 \caption{A typical time domain representation of a postmerger waveform 
 (THC:0001~\cite{Dietrich:2018phi,Radice:2016rys,Radice:2017lry}); 
 cf.~Tab.~\ref{tab:NR_data}.
 As throughout the article, we restrict our considerations to the 
 dominant 22-mode.}
 \label{fig:qualitative_TD}
\end{figure}

While the inspiral GW signal is characterized by a chirp, 
i.e., a monotonic increase of the GW amplitude and frequency, 
the postmerger emission shows a non-monotonic amplitude and 
frequency evolution. 
Figure~\ref{fig:qualitative_TD} presents one example of a possible postmerger waveform. 
In the following, we highlight some of the important 
features characterizing the signal. \\

\textbf{First postmerger minimum:}
By definition, the inspiral ends at the peak of the GW amplitude (merger) 
marked with a red circle in Fig.~\ref{fig:qualitative_TD}.
After the merger, the amplitude decreases showing a clear minimum (red squared marker)
shortly afterwards, see~\cite{Thierfelder:2011yi,Kastaun:2016elu} for further discussions. 
Around this intermediate and highly non-linear regime, 
different frequencies are excited for a few milliseconds, 
see e.g.,~\cite{Bauswein:2015yca,Rezzolla:2016nxn} for further details. 
While it was already known that the merger frequency can be expressed by 
a quasi-universal relation, e.g.,~\cite{Takami:2014zpa,Bernuzzi:2014kca}, 
we find that the time between merger and this amplitude minimum 
also follows a similar relation. 

\begin{figure}[t]
 \centering
 \includegraphics[width=\columnwidth]{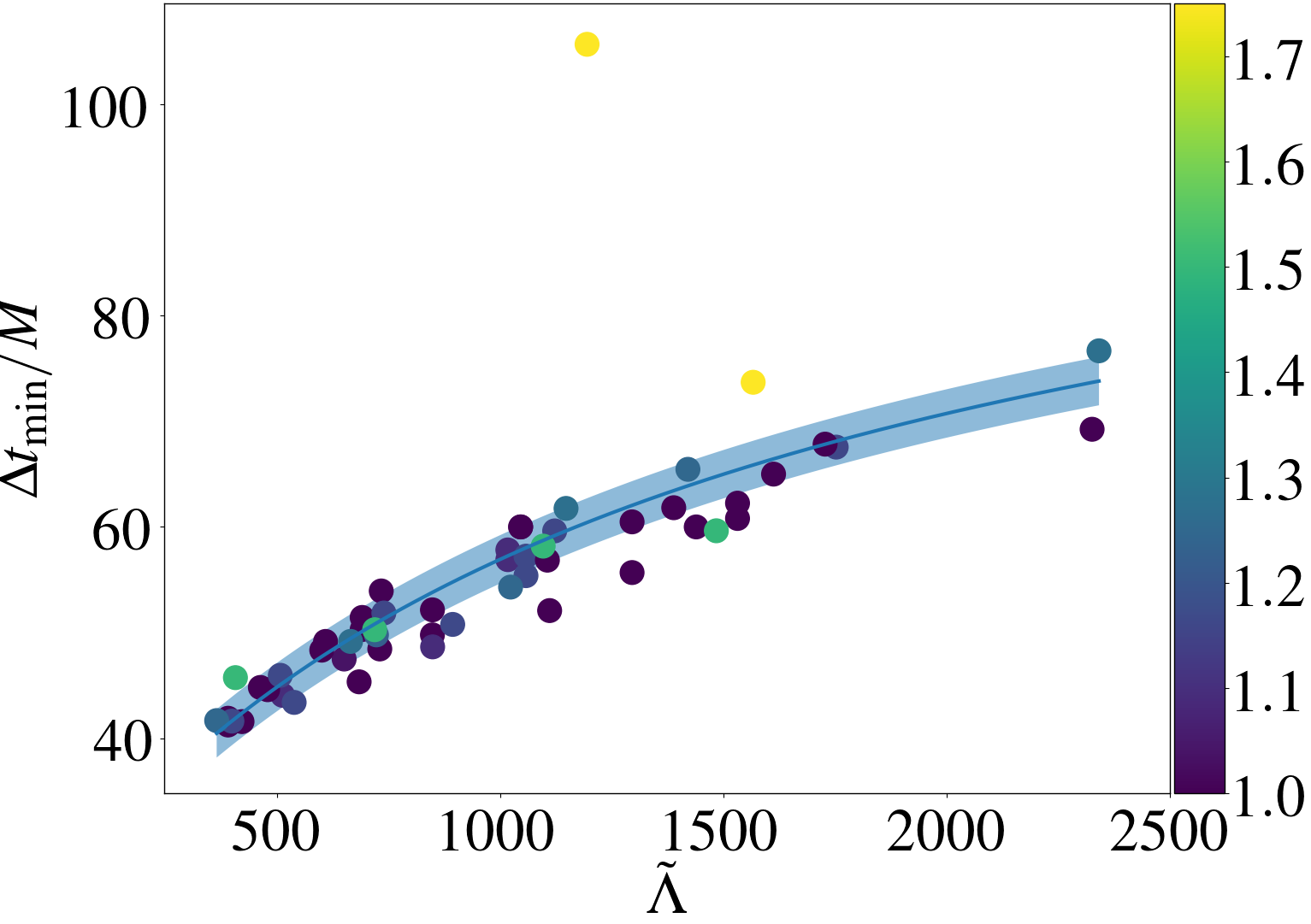}
 \caption{Dimensionless time between the merger and the 
          first amplitude minimum within the postmerger 
          as a function of the mass-weighted tidal deformability $\tilde{\Lambda}$.
          The color shows the mass ratio $q$.
          The shaded region indicates the 1-sigma ($\pm 2.2913$) uncertainty.
         }
 \label{fig:dt_min}
\end{figure}

In Fig.~\ref{fig:dt_min} we show the time 
between the merger and the first amplitude minimum, 
$\Delta t_{\rm min}/M$, as a function of the mass-weighted tidal deformability 
\begin{align}
\tilde{\Lambda} = \frac{16}{13} \frac{ (M_A+12M_B)M_A^4\Lambda_A 
+ (M_B+12M_A)M_B^4\Lambda_B }{ (M_A+M_B)^5 } \,. 
\label{eq:tildeLambda}
\end{align}
with the individual dimensionless tidal deformabilities 
$\Lambda = \frac{2}{3} k_2 (\frac{R}{M})^5$, 
where $k_2$ labels the dimensionless $\ell=2$ Love number and 
$R$ labels the radius of the isolated NSs. 
We show with different colors the 
mass ratio of each setup defined as $q=M_A/M_B \geqslant 1$, 
cf.~colorbar of Fig.~\ref{fig:dt_min}.

We find a clear correlation between $\Delta t_{\rm min}/M$ and 
the mass-weighted tidal deformability $\tilde{\Lambda}$. 
A good phenomenological representation is given by 
\begin{equation}
\frac{\Delta t_{\rm min}}{M} = \alpha \frac{1 + \beta \tilde{\Lambda}}{1+ \gamma \tilde{\Lambda}}, \label{eq:tmin_qu}
\end{equation}
with the parameters $\alpha=2.4681 \times 10^1, \beta=2.8477 \times 10^{-3}, 
\gamma = 6.6798 \times 10^{-4}$ obtained by a least-square fit for which the 
root-mean-square (RMS) error is $2.4608$. 
Interestingly, the two highest mass ratio simulations do not follow Eq.~\eqref{eq:tmin_qu}. 
This is caused by the different postmerger evolution for these high-mass ratio setups. 
While the amplitude minimum is produced when the two NS cores approach each other 
and potentially get repelled, configurations with very high mass ratio show almost a  
disruption during the merger, i.e., the lower massive NS deforms 
significantly under the strong external gravitational field 
of its companion. \\

One possible application for the quasi-universal relation for 
$\Delta t_{\rm min}/M$ is the improvement of BNS waveform approximants, i.e., 
it might help to determine the amplitude evolution after the merger of the two NSs. 
In particular, incorporating an amplitude tapering after the merger with a 
width of $\Delta t_{\rm min}$ provides a natural ending condition for inspiral-only 
approximants, e.g.,
\texttt{NRTidal}~\cite{Dietrich:2017aum,Dietrich:2018uni,Dietrich:2019kaq} 
or tidal effective-one-body models~\cite{Bernuzzi:2014owa,Hinderer:2016eia,Nagar:2018zoe}. 
Therefore, Eq.~\eqref{eq:tmin_qu} might become a central criterion to connect inspiral 
and postmerger models.\\

\begin{figure}[t]
 \centering
 \includegraphics[width=\columnwidth]{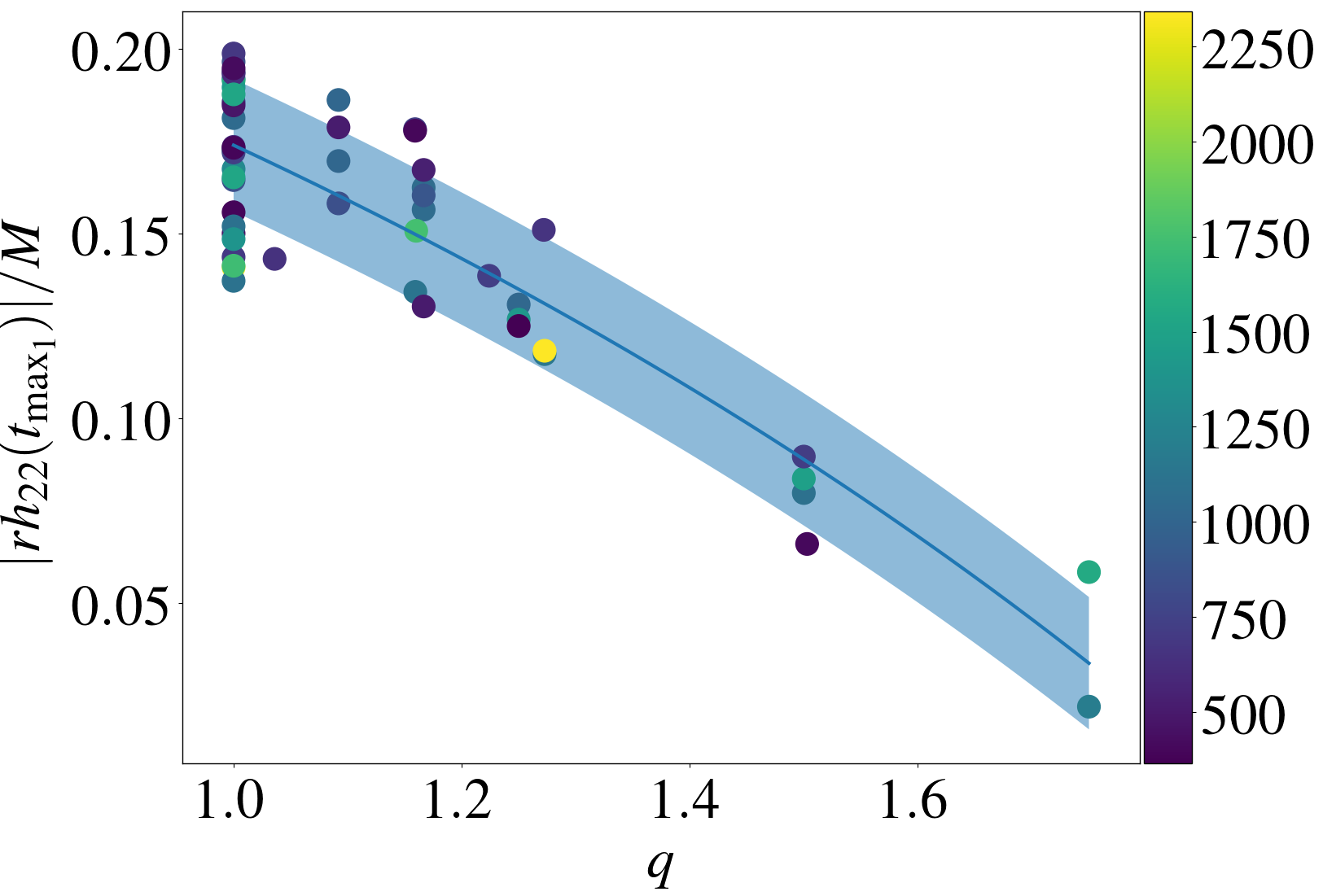}
 \caption{A scatter plot of the first peak in the postmerger spectra after merger versus $q$.
          The color shows the mass-weighted tidal deformability $\tilde{\Lambda}$.
          The shaded region indicates the 1-sigma ($\pm 1.7915 \times 10^{-2}$) uncertainty.
         }
 \label{fig:TD_firstpeak_toverM_q}
\end{figure}

\textbf{First postmerger maximum:}
After the minimum of the GW amplitude, the amplitude grows and reaches 
a maximum, marked with a red diamond in Fig.~\ref{fig:qualitative_TD}. 
One finds that the main binary property determining the amplitude of this first postmerger 
GW amplitude maximum is the mass ratio of the binary $q$, 
cf.~Fig.~\ref{fig:TD_firstpeak_toverM_q} with
\begin{equation}
 \dfrac{|rh_{\rm 22}(t_{\rm {max}_1})|}{M}
 = (2.8437 \times 10^{-1}) \frac{1-(5.3149 \times 10^{-1}) q}{1-(2.3420\times 10^{-1}) q}.
\end{equation}
The qualitative behavior is again related to the possible tidal disruption of 
the binary close to the merger for unequal mass systems. 
We note that even if the secondary star does not get disrupted, the maximum density in the remnant 
shows one peak rather than two independent cores~\cite{Kastaun:2016yaf,Hanauske:2019qgs} 
which leads to a smaller first postmerger peak and overall on average a smaller GW amplitude. 

\subsection{Frequency domain}

\begin{figure}[t]
 \centering
 \includegraphics[width=\columnwidth]{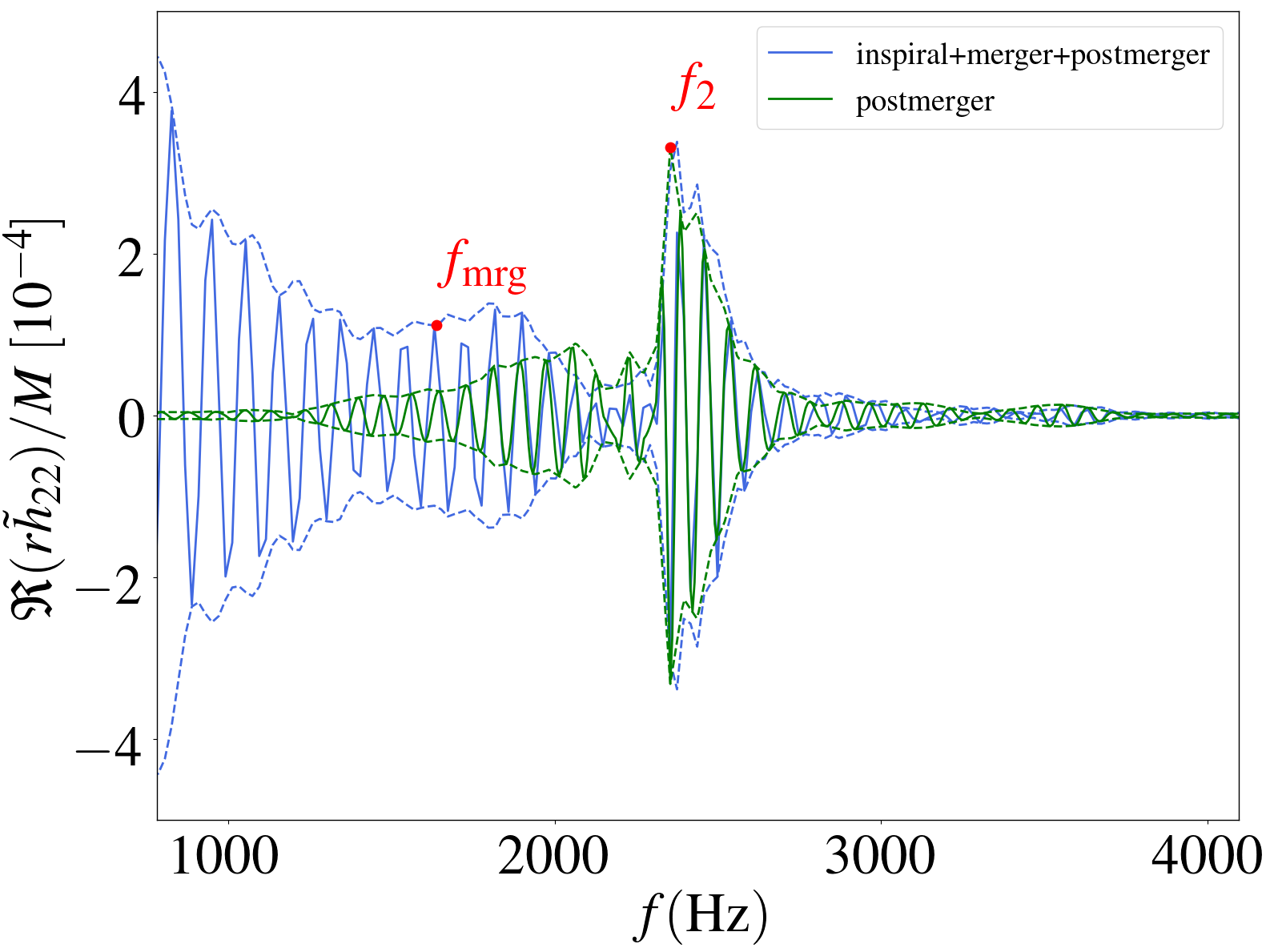}
 \caption{Frequency domain gravitational waveform (for the dominant 22 mode) for the setup 
 THC:0001. In blue we show the FFT of the time domain waveform shown in Fig.~\ref{fig:qualitative_TD}, 
 while in green we show only the postmerger part obtained by discarding the inspiral signal 
 in the time domain. The two marked characteristic frequencies are the merger frequency $f_{\rm mrg}$
 and the main postmerger emission frequency $f_2$.}
 \label{fig:qualitative_FD}
\end{figure}

We obtain the frequency domain waveform by fast Fourier transformation (FFT) of the 
time domain GW strain $h(t)$:
\begin{equation}
 \tilde{h}(f) = \int^{\infty}_{-\infty}  h(t) e^{-i 2 \pi f t} \dd t.
 \label{eq.FFT}
\end{equation}
As before, we consider only the dominant 22-mode of the GW signal. 

In Fig.~\ref{fig:qualitative_FD} we show the frequency domain GW signal (blue solid line)
of THC:0001. The merger frequency of this particular 
configuration is 1638 Hz and it is marked with a red circle. 
The main feature of the postmerger spectrum is 
the dominant peak characterizing the main emission frequency $f_2$, 
which for the setup shown in the Figs.~\ref{fig:qualitative_TD} 
and \ref{fig:qualitative_FD} is about 2354 Hz. 
For a better interpretation, we also present the frequency domain postmerger spectrum 
in green. Such postmerger-only waveforms are obtained by FFT
after applying a Tukey window~\cite{harris:1978} with a shape parameter 0.05 at $t_{\rm min}$ 
(where the shape parameter represents the fraction of the window inside the cosine tapered region)
and will be used for our injections to test our parameter estimation infrastructure. \\

\begin{figure}[t]
 \centering
 \includegraphics[width=\columnwidth]{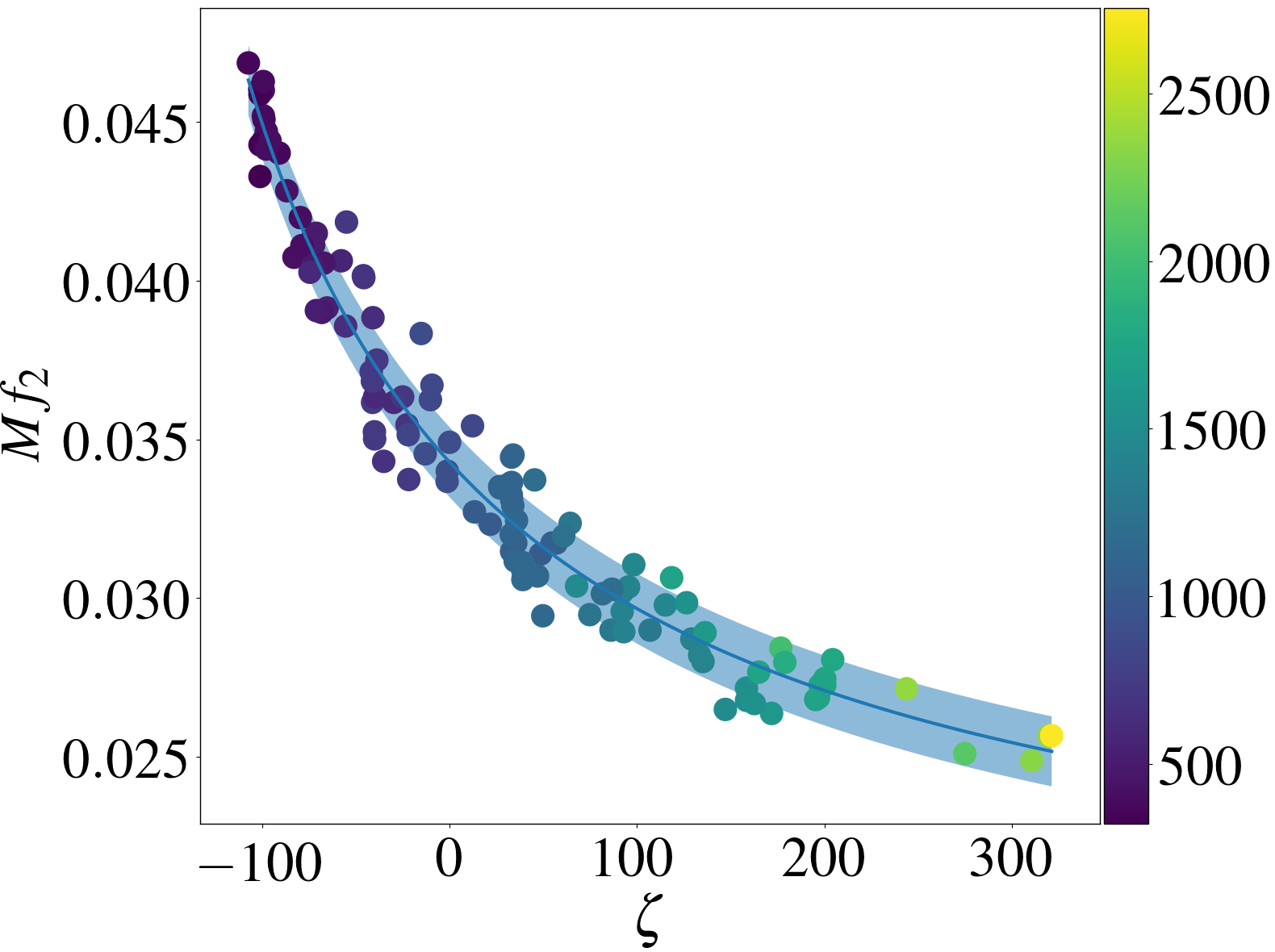}
 \caption{$Mf_2$ as a function of $\zeta$.
          The color shows the mass-weighted tidal deformability $\tilde{\Lambda}$.
          The shaded region indicates the 1-sigma ($\pm 1.1025\times 10^{-3}$) uncertainty.
          In addition to the CoRe-dataset employed to derive the previously shown 
          quasi-universal relations, we include here the published results 
          of~\cite{Takami:2014tva,Hotokezaka:2013iia}. 
         }
 \label{fig:f2_zeta}
\end{figure}

\textbf{$f_2$-frequency:}
The dominant feature in the postmerger frequency spectrum is the 
dominant emission mode of the merger remnant at a frequency $f_2$.
As mentioned before, a number of works, 
e.g.,~\cite{Bauswein:2011tp,Clark:2014wua,Takami:2014zpa,
Bernuzzi:2015rla,Rezzolla:2016nxn} have discussed possible EOS-insensitive, 
quasi-universal relation for the $f_2$-frequency. 


Building mostly on the work of Bernuzzi \emph{et al.}~\cite{Bernuzzi:2015rla}, 
we derive a new relation for the $f_2$ frequency. 
First, we extend the dataset of 99 NR simulations 
employed in Ref.~\cite{Bernuzzi:2015rla} and use a set of 121 data by 
incorporating additional setups published as a 
part of the CoRe database~\cite{Dietrich:2018phi}; cf.~Tab.~\ref{tab:NR_data}.
Second, we are switching from
\begin{equation}
\kappa^{\rm T}_2 \equiv
2\left[\dfrac{1}{q}\left(\dfrac{X_{A}}{C_{A}}\right)^{5}k^{A}_{2}+
q\left(\dfrac{X_{B}}{C_{B}}\right)^{5}k^{B}_{2}\right], 
\end{equation}
to 
\begin{small}
\begin{equation}
\label{eq:kappa}
\kappa^{\rm T}_{\rm eff} \equiv \frac{2}{13} \left[
\left(1+12\frac{X_B}{X_A}\right)\left(\frac{X_A}{C_A}\right)^5 k^A_2 +
 (A \leftrightarrow B) \right]  \ = \frac{3}{16} \tilde{\Lambda}
\end{equation}
\end{small}
which yields a tiny improvement ($\sim 0.1 \%$) in the RMS
error against the NR data, but more notably relates directly 
to the mass-weighted tidal deformability $\tilde{\Lambda}$ measured most accurately 
from the inspiral part of the signal. 
In addition to the dependence of the mass-weighted tidal deformability $\tilde{\Lambda}$, 
the postmerger evolution depends also on the stability of the formed remnant 
and how close it is to the black hole formation.
This information is in part encoded in the ratio between
the total mass $M$ and the maximum allowed mass of a single non-rotating NS $M_\mathrm{TOV}$. 
By incorporating an additional $M/M_{\rm TOV}$ dependence we are able to reduce 
the RMS error by $\approx 28\%$.

Therefore, we define a parameter $\zeta$ by a linear combination of 
$\kappa_\mathrm{eff}^{\rm T}$ and $\frac{M}{M_\mathrm{TOV}}$  
(see also~\cite{Coughlin:2018fis,Zappa_inprep,Zappa_dcc} for a similar approach), 
\begin{equation}
 \zeta = \kappa_\mathrm{eff}^{\rm T} + a \dfrac{M}{M_\mathrm{TOV}}.
 \label{eq:zeta}
\end{equation}
The free parameter $a=-131.7010$ is determined by minimizing the RMS error.
Finally, the dimensionless frequency $Mf_2$ is fitted against $\zeta$ 
using a Pad\'e approximant:
\begin{equation}
 Mf_2(\zeta) = \alpha \dfrac{1+A \zeta}{1+B \zeta}
 \label{eq:pade22}
\end{equation}
with $\alpha=3.4285 \times 10^{-2}$, $A=2.0796 \times 10^{-3}$ and $B=3.9588 \times 10^{-3}$. 
We present Eq.~\eqref{eq:pade22} together with our NR
dataset and a one sigma uncertainty region (shaded area) 
in Fig.~\ref{fig:f2_zeta}. \\


\begin{figure}[t]
 \centering
 \includegraphics[width=\columnwidth]{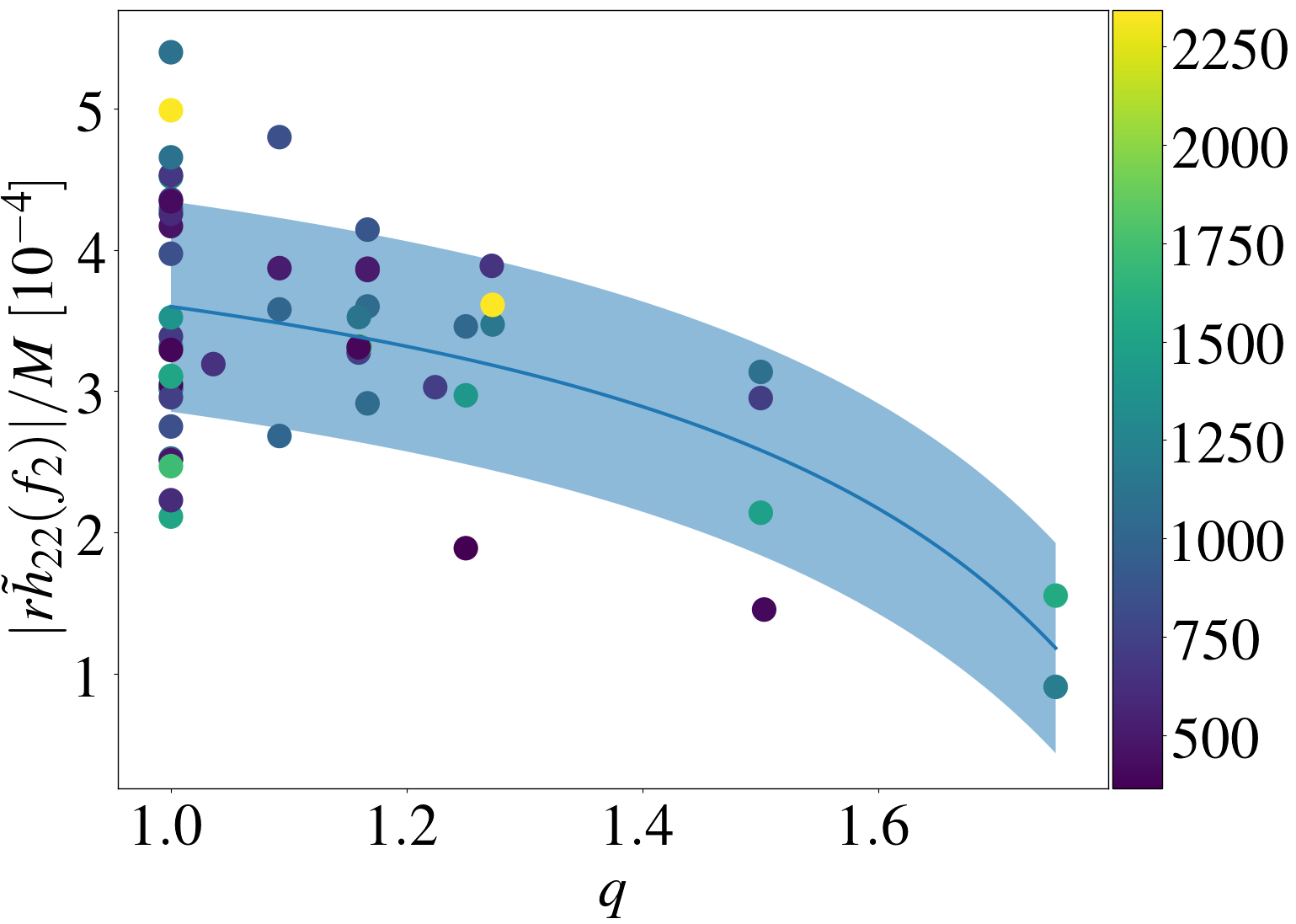}
 \caption{$|r\tilde{h_{\rm 22}}(f_2)/M|$ as a function of the mass ratio $q$ with 1-sigma 
          ($\pm7.4485\times10^{-5}$) shaded region.
          The color shows the tidal deformability $\tilde{\Lambda}$.
         }
 \label{fig:f2Amp_q}
\end{figure}

\textbf{Frequency domain amplitude of $f_2$:}
Finally, we want to briefly discuss the dependence of 
the $f_2$-peak amplitude on the binary properties. 
While the $f_2$-frequency correlates clearly to $\zeta$, 
we have not been able to find a similar tight relation between 
any combination of the binary parameters and the amplitude $|\tilde{h}_{\rm 22}(f_2)|$. 
The only noticeably imprint which we have been able to extract comes from 
the mass ratio $q$, where generally higher mass ratios lead to a smaller 
amplitude $|r\tilde{h}_{\rm 22}(f_2)/M|$ as shown in Fig.~\ref{fig:f2Amp_q} with 
\begin{equation}
 \dfrac{|r\tilde{h}_{\rm 22}(f_2)|}{M}
 = (4.2319 \times 10^{-4}) \frac{1-(5.4016\times10^{-1}) q}{1-(4.5927\times10^{-1}) q}.
\label{eq:FD_f2_amp}
\end{equation}
We note that because of the large uncertainty, 
we see Eq.~\eqref{eq:FD_f2_amp} more as a qualitative rather 
than a quantitative statement about the postmerger spectrum.  
However, the overall amplitude decreases for an increasing mass ratio 
seems to be a robust feature and might help 
to interpret future GW observations.  

\section{Model functions, $f_2$ measurement, 
and inspiral-postmerger consistency}

\subsection{Lorentzian Approximants}
\label{sec:model_function}

Based on our previous discussion and the dominance of the characteristic $f_2$ frequency, 
we start our consideration with a simple damped sinusoidal time domain waveform 
to model the postmerger waveform. 
The Fourier transform of a damped sinusoidal function 
is a Lorentzian function, Eq.~\eqref{eq:Lorentzian}.
In the simplest case which we consider, we use 3 unknown coefficients
($c_0,c_1,c_2$) corresponding to the 
amplitude, the dominant emission frequency and the inverse of the damping time, respectively, 
and write the frequency-domain signal as: 
\begin{equation}
 \tilde{h}_{22}(f) = \dfrac{c_0\ c_2}{\sqrt{(f-c_1)^2+ c_2^2}} e^{-i\arctan(\frac{f-c_1}{c_2})}.
 \label{eq:Lorentzian}
\end{equation}
Equation~\eqref{eq:Lorentzian} suggests that the amplitude peak of the GW postmerger 
spectrum and also the main postmerger phase evolution are connected
to the same frequency characterized by $c_1$. 

\begin{figure}[t]
 \centering
 \includegraphics[width=\columnwidth]{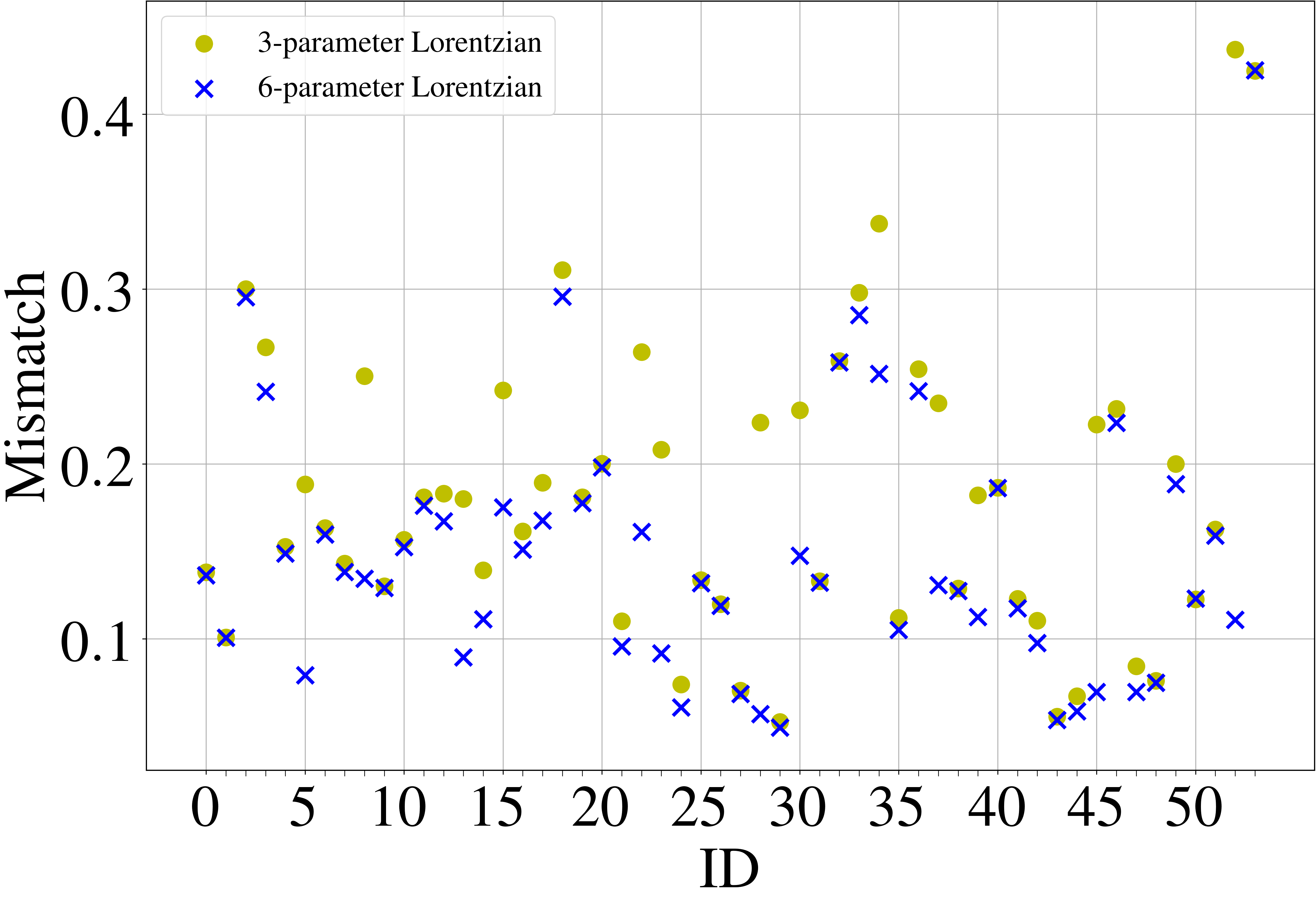}
 \caption{Mismatch between a subset of the NR data listed in Tab.~\ref{tab:NR_data} 
          and the three- and six-parameter models. 
         }
 \label{fig:mismatch_diff_approx}
\end{figure}

Maximizing over ($c_0,c_1,c_2$), we compute the mismatches between 
the used NR data from the CoRe-database (Tab.~\ref{tab:NR_data}) 
and the model function, Eq.~\eqref{eq:Lorentzian}. 
Figure~\ref{fig:mismatch_diff_approx} shows all mismatches, which on average 
are $\sim 0.18$. 

The mismatches can be further decreased by adding 
three additional coefficients:
\begin{equation}
 \tilde{h}_{22}(f) = \dfrac{c_0\ c_2}{\sqrt{(f-c_1)^2+ c_2^2}} e^{-ic_3 \arctan(\frac{f-c_5}{c_4})}
 \label{eq:6paramsLorentzian}.
\end{equation} 
For Eq.~\eqref{eq:6paramsLorentzian} the amplitude and phase 
evolution are independent from each other
and we obtain average mismatches of $0.15$, i.e., 
about $17\%$ better than for the three-parameter model. 
While one might argue that the additional introduced degrees of freedom 
hinder the extraction of individual parameters in a full Bayesian analysis, 
it might also be possible that the more flexible 6-parameter model
recovers signals with smaller SNRs. 
Thus, we continue our study with both model functions 
Eqs.~\eqref{eq:Lorentzian} and \eqref{eq:6paramsLorentzian}.

Finally, one obtains the plus and cross polarizations from Eq.~\eqref{eq:Lorentzian} and 
Eq.~\eqref{eq:6paramsLorentzian} by incorporating the inclination ($\iota$)
dependence:
\begin{align}
 \tilde{h}_p &= \dfrac{1+\cos^2(\iota)}{2} \tilde{h}_{22}, \\
 \tilde{h}_c &= -i\cos(\iota) \tilde{h}_{22}.
 \label{eq:hphc}
\end{align}

$\tilde{h}_c,\tilde{h}_p$ can be employed directly to infer information 
from the postmerger part of a GW signal or to construct a 
full IMP-waveform for BNSs. 

\subsection{Validating the Parameter Estimation Pipeline}
\label{sec:parameter_estimation}

\begin{figure}[t]
 \centering
 \includegraphics[width=0.98 \columnwidth]{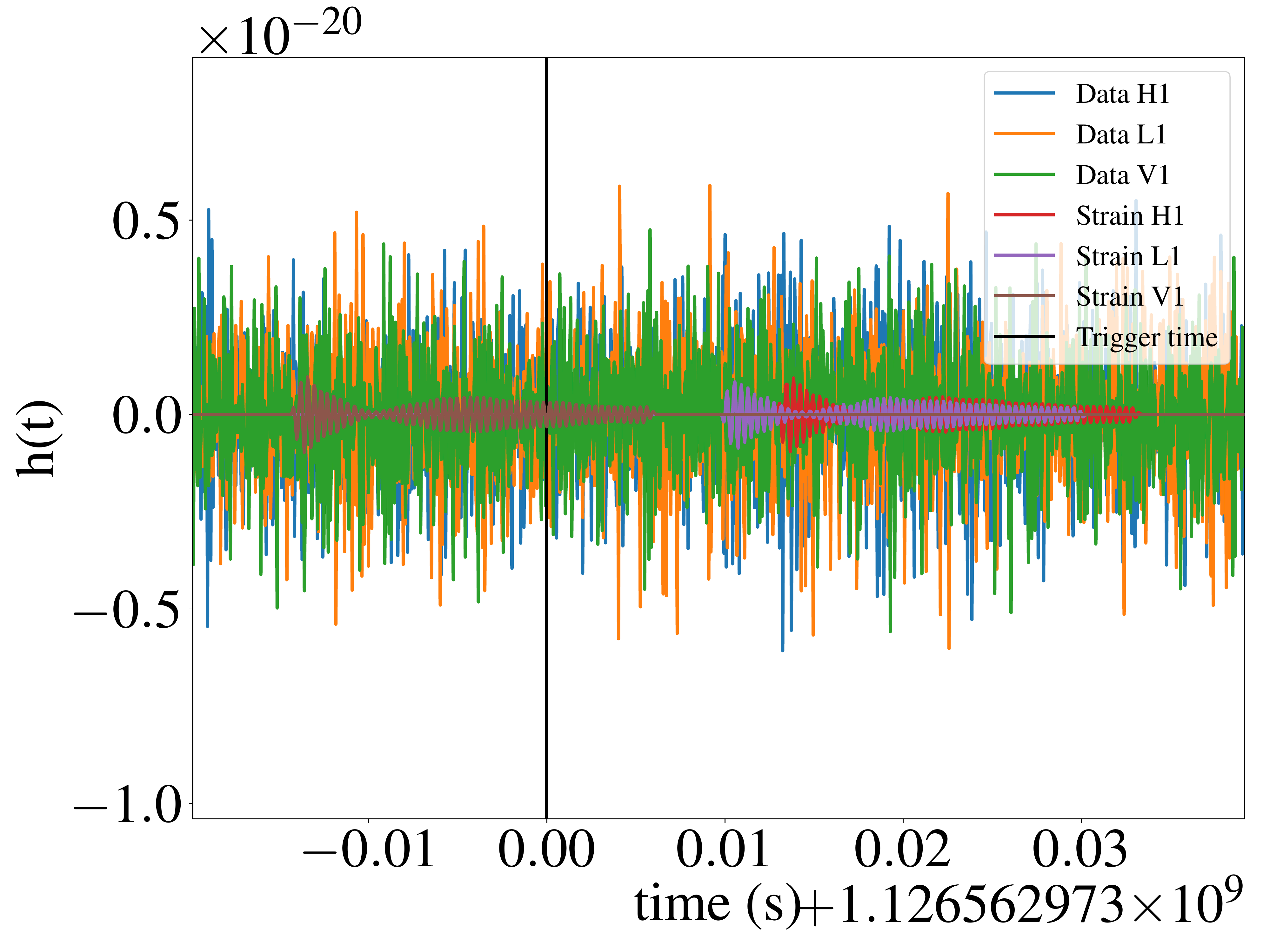} \\
 \includegraphics[width=0.98 \columnwidth]{fig08b.pdf} 
 \caption{Injection of THC:0021 with SNR 8, zero inclination angle,
 zero polarization angle and sky location (0,0).
 Top panel: time domain signal of the postmerger waveform
 highlighted within the detectors noise for H1--Hanford, L1--Livingston, and V1--Virgo. 
 Bottom panel: frequency domain signal and design amplitude spectral density (ASD) for H1, L1, and V1.
 }
 \label{fig:THC0021_SNR8}
\end{figure}

In this section, we present for four selected cases 
the performance of the three- and six-parameter models.
We inject the NR waveforms immersed in the same simulated Gaussian noise
with total network SNRs ranging from SNR 0 to SNR 10
assuming that Advanced LIGO and Advanced Virgo detectors
run at design sensitivity~\cite{Aasi:2013wya}~\footnote{
The corresponding power spectral density (PSD) files,
\texttt{LIGO-P1200087-v18-aLIGO\_DESIGN.txt} and
\texttt{LIGO-P1200087-v18-AdV\_DESIGN.txt}
are available under the \texttt{LALSimulation} module in the \texttt{LALSuite} package.}. 
Fig.~\ref{fig:THC0021_SNR8} shows the injection of THC:0021 with SNR 8 in both the time (top panel) 
and frequency domain (bottom panel).
For each injected waveform, a Tukey window with shape parameter 0.05 is applied at $t_{\rm min}$ 
to isolate the postmerger signal and avoid Gibbs phenomenon.
All simulated signals are injected with zero inclination angle,
zero polarization angle $\psi$ and sky location ($\alpha, \delta$) to be ($0, 0$).

We estimate parameters using Bayesian inference
with the \texttt{LALInference} module~\cite{Veitch:2014wba}
available in the \texttt{LALSuite} package.
Sampling is done on 9 (12) parameters
\begin{equation}
 \{ c_i, \alpha, \delta, \iota, \psi, t_c, \phi_c \}
 \label{samplingparams}
\end{equation}
with nested sampling algorithm \texttt{lalinferencenest}~\cite{Veitch:2009hd},
where $i$ runs from 0 to 2 (5) for the three- (six-) parameter model, and
$t_c$ and $\phi_c$ are the reference time and phase, respectively.
The priors are chosen to be uniform in $[0, 10^{-20}] {\rm s^{-1}}$ on $c_0$,
uniform in $[1500, 4096] {\rm Hz}$ on $c_1$ and $c_5$, 
uniform in $[1, 400] {\rm Hz}$ on $c_2$ and $c_4$,
uniform in $[0, 6]$ on $c_3$,
uniform in $[0, 2\pi]$ on $\alpha, \psi$ and $\phi_c$,
uniform in $[-1, 1]$ on $\cos(\iota)$ and $\sin(\delta)$, and
uniform in $[{\rm trigger~time} - 0.05 {\rm s}, {\rm trigger~time} + 0.05 {\rm s}]$ on $t_c$
where the trigger time is the signal arrival time at the geocentric frame.

\begin{figure*}[t]
 \centering
 \includegraphics[width=0.98 \textwidth]{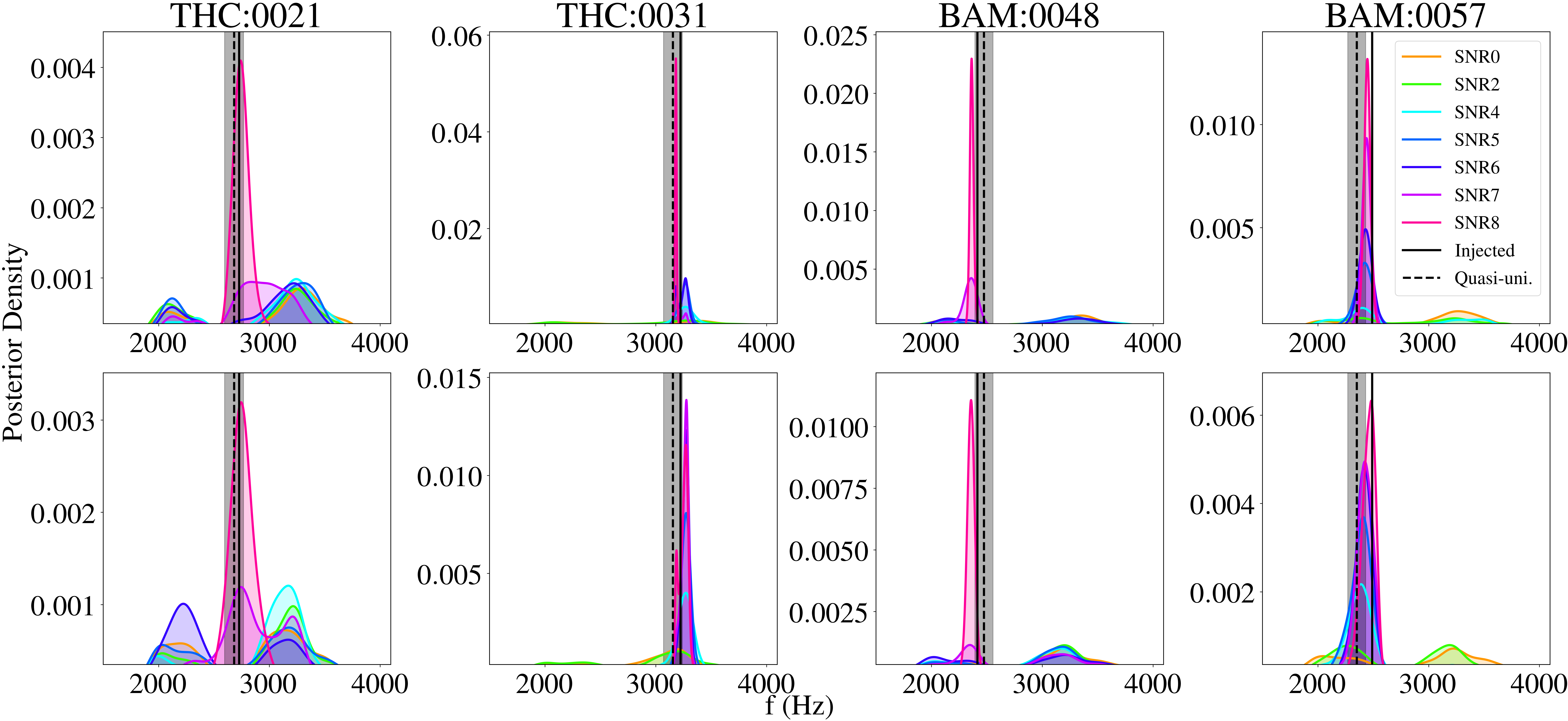} 
 \caption{Posteriors for the parameter $c_1$ in Eq.~\eqref{eq:Lorentzian} (top panels)
 and Eq.~\eqref{eq:6paramsLorentzian} (bottom panels) for a variety of SNRs. 
 $c_1$ can be directly related to the peak in the frequency domain spectrum and 
 therefore relates to the $f_2$ frequency. 
 The IDs of the four injected waveforms are \# 18, 20, 32, 34, i.e., 
 THC:0021, THC:0031, BAM:0048, BAM:0057 of~\cite{Dietrich:2018phi}. 
 The chosen set covers various EOSs, mass ratios, and masses and is 
 therefore used as a testbed for our new algorithm.}
 \label{fig:3and6params_dchi1}
\end{figure*}

Fig.~\ref{fig:3and6params_dchi1} shows the posterior for $c_1$, i.e., our best estimate 
of the $f_2$-frequency for SNRs up to 8 for our 4 examples, 
which we mark in Tab.~\ref{tab:NR_data}. 
We present the recovery with the 3- and the 6-parameter model 
in the top and bottom panels, respectively. 
The solid vertical line represents the injected $f_2$-frequency and 
the dashed line represents the estimate according to the quasi-universal 
relation Eq.~\eqref{eq:pade22} together with a one-sigma uncertainty (gray shaded region). 

\begin{figure}[t]
 \centering
 \includegraphics[width=0.98\columnwidth]{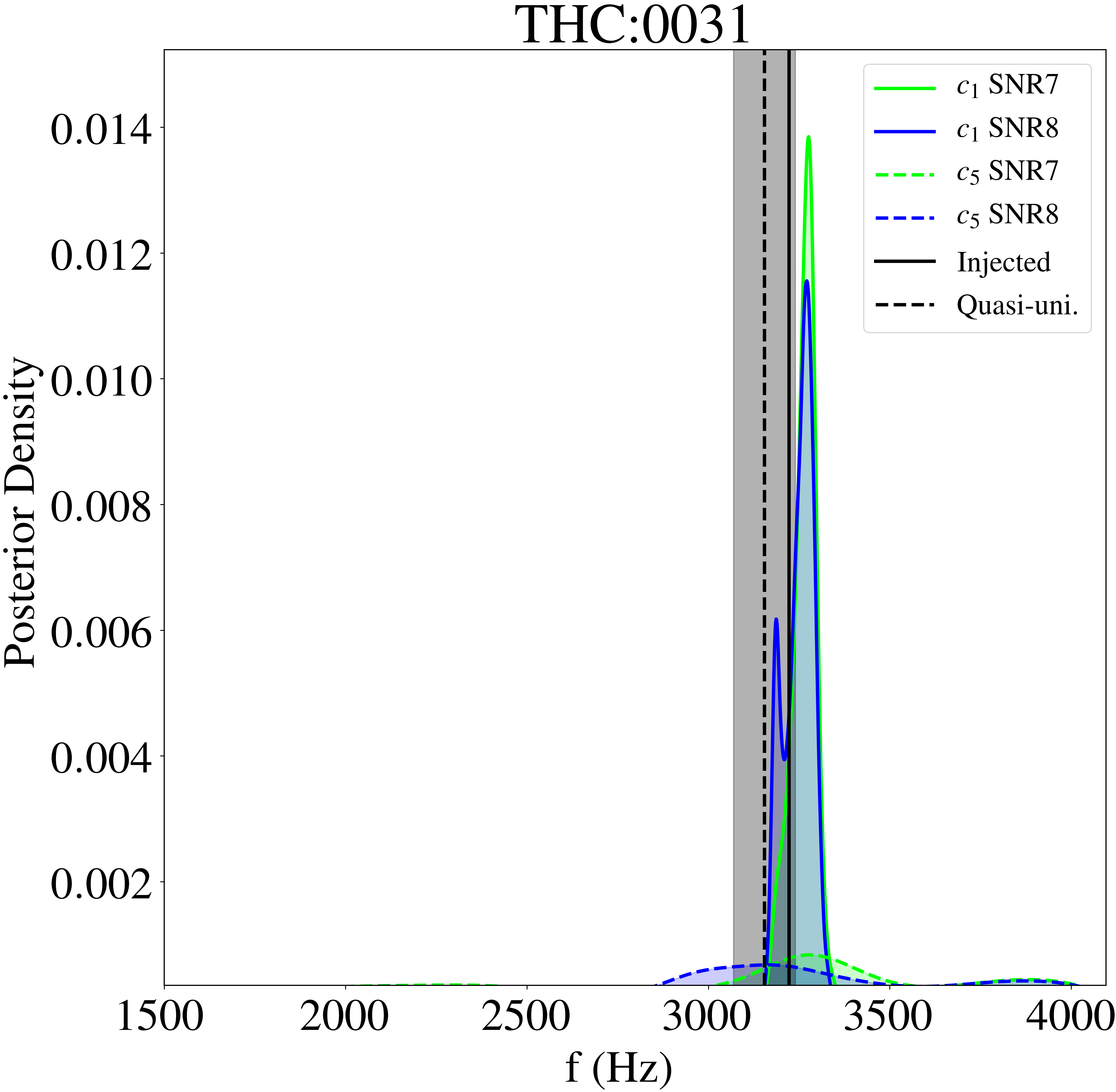}
 \caption{Posteriors for the parameter $c_1$ and $c_5$ in Eq.~\eqref{eq:6paramsLorentzian}
          for 2 SNRs. $c_5$ peaks at a frequency close to $f_2$ but it 
          is significantly less constrained than $c_1$.
         }
 \label{fig:c1c5_posterior_THC0031}
\end{figure}

We summarize the main findings below: 
\begin{enumerate}[(i)]
 \item 
 The three-parameter and six-parameter approximants 
 perform similarly.
 \item Depending on the exact setting (e.g., intrinsic source properties, noise realization, 
 sky location) one can recover the $f_2$ frequency with an SNR of $\sim 4$ for the best 
 and $\sim 8$ for worst considered scenarios.
 \item Interestingly, one finds that also $c_5$ relates to a frequency 
 which is close to the $f_2$ frequency, however, $c_5$ 
 is significantly less constrained than $c_1$ 
 (Fig.~\ref{fig:c1c5_posterior_THC0031}).
 \item Once 3rd generation detectors are available 
 and so the postmerger SNRs of $\sim 10$ are obtained, 
 the systematic uncertainties of the quasi-universal 
 relations become larger than the statistical uncertainties; 
 cf.~dashed and solid, vertical black lines. 
\end{enumerate}

\subsection{Inspiral and Postmerger consistency}
\label{sec:results}

\begin{figure*}[t]
 \centering
 \includegraphics[width=\textwidth]{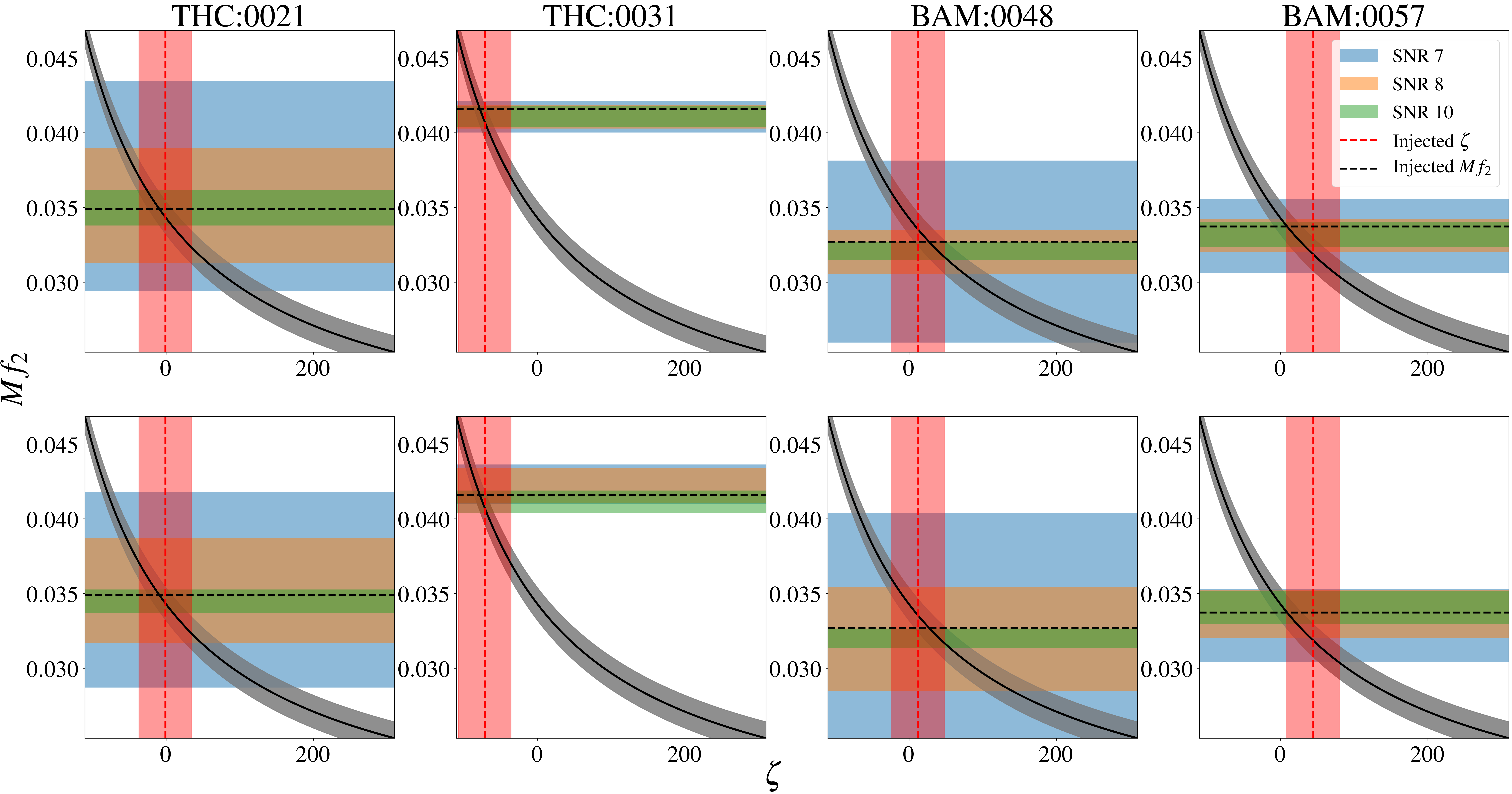}
 \caption{Schematic plot showing how one can constrain $\zeta$ from $f_2$ measurements
          where the spread $f_2$ are measured by one standard deviation.
          Top panel is for three-parameter model and the bottom panel 
          is for six-parameter model.
          The vertical red shaded region corresponds to the $\zeta$-interval consistent 
          with a hypothetical inspiral signal assuming an uncertainty of $\pm 0.04M_\odot$ 
          for $M$ and $M_{\rm TOV}$, and $\pm 30$ for $\kappa_{\rm eff}^{\rm T}$.
          The exact value is marked as vertical red-dashed line.
         }
 \label{fig:bounds_plot}
\end{figure*}

Finally, we want to illustrate how a future detection of a postmerger GW signal 
will help to constrain the source properties and the internal composition of NSs. 
As shown before, the $f_2$-frequency can be extracted through a 
simple waveform model (or alternatively by using \texttt{BayesWave}, 
e.g.,~\cite{Chatziioannou:2017ixj}). 
To connect the $f_2$-frequency with the source parameters, 
one needs to employ quasi-universal relations as presented 
in Sec.~\ref{sec:description} and some information obtained 
from the analysis of the inspiral GW signal. 
In particular, the total mass $M$ can be measured precisely using
state-of-art BNS inspiral waveforms, 
e.g.,~\cite{Damour:2012yf,Dietrich:2018uni,Kawaguchi:2018gvj,Messina:2019uby,
Schmidt:2019wrl,Dietrich:2019kaq,Bernuzzi:2014owa,Hinderer:2016eia,
Nagar:2018zoe,Lackey:2018zvw,Lange:2017wki,Lange:2018pyp}. 
For GW170817 the uncertainty of $M$ could be reduced to $\pm 0.04M_\odot$ once 
EM information had been included~\cite{Coughlin:2018fis,Radice:2018ozg}. Thus, 
we will use an uncertainty of $\Delta M = \pm 0.04M_\odot$ as a conservative estimate. 
In addition, we have to know the maximum TOV-mass $M_{\rm TOV}$. 
Current estimates for $M_{\rm TOV}$ are based on the observation of 
J0740+6620~\citep{Cromartie:2019kug} with $M= 2.17^{+0.11}_{-0.10} M_\odot$ and 
the assumption that GW170817's endstate was a black 
hole~\cite{Margalit:2017dij,Rezzolla:2017aly,Ruiz:2017due,Shibata:2019ctb} 
such that $M_{\rm TOV} \lesssim 2.17 $-$2.35 M_\odot$. 
Due to the increasing number of BNS detections in the future, 
we expect that the uncertainty of $M_{\rm TOV}$ can be considerably reduced 
so that we will use an uncertainty of $\pm 0.04M_\odot$. 
From this information, we can compute the 
$\zeta$-interval consistent with the observed inspiral signal 
from the $\tilde{\Lambda}$ posteriors, 
cf.~vertical red shaded region in Fig.~\ref{fig:bounds_plot}. 
We then connect the $\zeta$ estimate obtained from the inspiral 
with Eq.~\eqref{eq:pade22} and the $f_2$-measurement of the postmerger signal.
This consistency analysis is somewhat connected to the inspiral-merger-ringdown consistency 
test for BBH~\cite{Ghosh:2017gfp}, 
but not only has to assume the correctness of general relativity but also that 
our understanding of supranuclear matter and the EOS-insensitive 
quasi-universal relations are valid. 

Figure~\ref{fig:bounds_plot} summarizes our main results.
Generally the GW measurements can be considered consistent between the inspiral and postmerger observations 
as well as with the quasi-universal relation relating the main postmerger frequency with the binary properties. 
We find that for all cases, the quasi-universal relation and its 1-sigma uncertainty region lie within the 
intersection of the red shaded region (inspiral) and the blue/orange/green horizontal regions (postmerger).
Thus, (as expected) all simulations are consistent with (i) general relativity and (ii) the nuclear physics 
descriptions used as a basis for NR simulations to derive the quasi-universal relations. 
For future events this approach will allow us to probe our understanding of physical processes under 
extreme conditions, and in cases where the quasi-universal relation seems violated to even 
derive new relations based on GW measurements (under the assumption that general relativity is correct). 
We note that even in the case where either general relativity or the quasi-universal relations would be violated, 
we might not be able to determine reliably the violation based on one individual event,
but stronger constraints can be obtained by combining multiple BNS events.

\section{Summary}
\label{sec:conclusions}

In this work, we discussed the general morphology of a BNS postmerger in both the time and 
frequency domain.
We presented quasi-universal relations for the time at 
which the first postmerger amplitude minimum happens and 
the strength of the first postmerger amplitude maximum. 
In general, the time between the merger and the amplitude minimum increases 
with an increasing $\tilde{\Lambda}$, 
while the amplitude of the first postmerger maximum 
decreases with an increasing mass ratio.  
In the frequency domain, we improved the existing quasi-universal relations of $Mf_2$
by extending the employed NR dataset (121 simulations in total) and 
adding an extra dependence of $M/M_{\rm TOV}$. 
The extra term $M/M_{\rm TOV}$ characterizes how close the setup is 
to the black hole formation.

We find that a three- (six-) parameter Lorentzian can model the
postmerger waveform with average mismatch of 0.18 (0.15).
To test these model functions, we performed an injection 
study, in which we simulated the detector strains
with four different BNS configurations immersed in the same simulated Gaussian noise
assuming Advanced LIGO and Advanced Virgo at design sensitivities.
We find that in the best cases the Lorentzian models
could measure the dominant emission frequency $f_2$ once the signal has an 
SNR of 4 or above; however, for most scenarios higher SNRs $\sim 8$ were required. 

Employing the new quasi-universal relation for $Mf_2$ described in this work,
we could present consistency tests between the inspiral and the postmerger signal; 
cf.~Fig.~\ref{fig:bounds_plot}.


\section*{Acknowledgments}

  We thank Anuradha Samajdar for support throughout the 
  project and Frank Ohme for a number of helpful comments. 
  We are also thankful for members of the Computational Relativity 
  Collaboration for discussions and input. 
  
  T.~D.~acknowledges support by the European Union's Horizon
  2020 research and innovation program under grant
  agreement No 749145, BNSmergers. 
  K.~W.~T., T.~D., and C.~v.~d.~B.\ are supported by the research programme
  of the Netherlands Organisation for Scientific Research (NWO).

\bibliographystyle{apsrev}
\bibliography{paper.bib}

\onecolumngrid
\clearpage
\appendix
\section{NR configurations}

\begin{center}
\begin{longtable}{cccccccccccc}
\toprule
\hline \hline
\vspace{-0.3cm}
&&&&&&&&&& \\
ID &      Code/CoRe-ID &     EOS &  $\mathrm{M}_\mathrm{TOV}$ [$\mathrm{M_\odot}$]  &    M [$\mathrm{M_\odot}$] &    q &  
$\kappa_\mathrm{eff}^{\rm T}$ &  $\dfrac{\Delta t_\mathrm{min}}{M}$ &  $\dfrac{|rh_{22}(t_\mathrm{{max}_1})|}{M} [10^{-2}]$  &      
$f_2 (\mathrm{Hz})$ &  $\dfrac{|r\tilde{h}_{22}(f_2)|}{M} [10^{-4}]$ \\
\vspace{-0.3cm}
&&&&&&&&&& \\
\hline
\midrule
\endhead
\midrule
\multicolumn{12}{r}{{Continued on next page}} \\
\midrule
\endfoot
\bottomrule
\hline \hline \\
\caption{All NR configurations employed to derive the quasi-universal relations. 
It includes simulations of the CoRe database~\cite{Dietrich:2018phi} labeled as `THC' or `BAM', 
and additionally results published in~\cite{Takami:2014tva} labeled as `whisky' and 
\cite{Hotokezaka:2013iia} labeled as `sacra'. 
We highlight the simulations which have been used for the injection study discussed in the main text.
The individual columns refer to: the number of simulation, the employed code, the EOS, 
the maximum TOV mass for the employed EOS, the total mass of the system, the mass ratio, 
the effective tidal coupling constant, the time at which the first postmerger minimum appears, 
the amplitude of the first postmerger maximum, the $f_2$ frequency, 
and the amplitude of the $f_2$ frequency peak. \label{tab:NR_data}} \\
\endlastfoot
\#0   &  THC:0001 &   BHBlp &   2.10 & 2.50 & 1.00 &       242 &                55.68 &             17.27 & 2354 &   3.32 \\
\#1   &  THC:0002 &   BHBlp &   2.10 & 2.60 & 1.00 &       196 &                60.00 &             19.20 & 2458 &   4.29 \\
\#2   &  THC:0003 &   BHBlp &   2.10 & 2.70 & 1.00 &       159 &                49.78 &             19.65 & 2726 &   3.97 \\
\#3   &  THC:0004 &   BHBlp &   2.10 & 2.62 & 1.09 &       190 &                56.90 &             18.62 & 2602 &   3.58 \\
\#4   &  THC:0005 &   BHBlp &   2.10 & 2.60 & 1.17 &       198 &                55.38 &             16.24 & 2478 &   3.60 \\
\#5   &  THC:0006 &   BHBlp &   2.10 & 2.80 & 1.00 &       129 &                51.43 &             18.57 & 2912 &   3.01 \\
\#6   &  THC:0007 &   BHBlp &   2.10 & 2.83 & 1.04 &       121 &                47.49 &             14.32 & 2767 &   3.19 \\
\#7   &  THC:0010 &     DD2 &   2.42 & 2.40 & 1.00 &       302 &                65.00 &             19.15 & 2231 &   3.11 \\
\#8   &  THC:0011 &     DD2 &   2.42 & 2.50 & 1.00 &       242 &                60.48 &             16.75 & 2354 &   3.02 \\
\#9   &  THC:0012 &     DD2 &   2.42 & 2.60 & 1.00 &       196 &                60.00 &             18.13 & 2478 &   2.52 \\
\#10  &  THC:0013 &     DD2 &   2.42 & 2.70 & 1.00 &       159 &                52.15 &             16.45 & 2664 &   2.75 \\
\#11  &  THC:0014 &     DD2 &   2.42 & 2.62 & 1.09 &       190 &                57.82 &             16.97 & 2437 &   2.68 \\
\#12  &  THC:0015 &     DD2 &   2.42 & 2.60 & 1.17 &       198 &                57.23 &             15.65 & 2478 &   2.91 \\
\#13  &  THC:0016 &     DD2 &   2.42 & 2.80 & 1.00 &       129 &                50.29 &             14.37 & 2571 &   3.39 \\
\#14  &  THC:0017 &     DD2 &   2.42 & 3.00 & 1.00 &        86 &                44.80 &             15.01 & 2767 &   4.17 \\
\#15  &  THC:0018 &   LS220 &   2.04 & 2.40 & 1.00 &       269 &                60.00 &             18.96 & 2520 &   4.52 \\
\#16  &  THC:0019 &   LS220 &   2.04 & 2.70 & 1.00 &       128 &                45.33 &             19.88 & 3015 &   4.53 \\
\#17  &  THC:0020 &   LS220 &   2.04 & 2.62 & 1.09 &       159 &                48.64 &             15.82 & 2850 &   4.80 \\
\rowcolor{gray!25}\#18  &  THC:0021 &   LS220 &   2.04 & 2.60 & 1.17 &       167 &                50.77 &             16.04 & 2726 &   4.14 \\
\#19  &  THC:0029 &    MS1b &   2.76 & 2.70 & 1.00 &       287 &                60.80 &             16.53 & 2014 &   2.11 \\
\rowcolor{gray!25}\#20  &  THC:0031 &    SFHo &   2.06 & 2.62 & 1.09 &        96 &                44.05 &             17.88 & 3222 &   3.87 \\
\#21  &  THC:0032 &    SFHo &   2.06 & 2.60 & 1.17 &       100 &                43.38 &             16.73 & 3056 &   3.86 \\
\#22  &  THC:0036 &     SLy &   2.06 & 2.70 & 1.00 &        73 &                41.24 &             15.58 & 3459 &   3.05 \\
\#23  &  BAM:0002 &      2H &   2.83 & 2.70 & 1.00 &       436 &                69.25 &             14.09 & 1871 &   4.99 \\
\#24  &  BAM:0003 &    ALF2 &   1.99 & 2.70 & 1.00 &       137 &                53.94 &             17.20 & 2720 &   2.96 \\
\#25  &  BAM:0004 &    ALF2 &   1.99 & 2.70 & 1.00 &       136 &                48.45 &             19.38 & 2791 &   4.36 \\
\#26  &  BAM:0009 &    ALF2 &   1.99 & 2.50 & 1.27 &       215 &                61.75 &             11.74 & 2391 &   3.47 \\
\#27  &  BAM:0010 &    ALF2 &   1.99 & 2.70 & 1.16 &       138 &                51.85 &             17.83 & 2633 &   3.27 \\
\#28  &  BAM:0022 &     ENG &   2.25 & 2.70 & 1.00 &        89 &                44.59 &             18.48 & 2933 &   2.51 \\
\#29  &  BAM:0035 &      H4 &   2.03 & 2.70 & 1.00 &       208 &                52.08 &             13.72 & 2406 &   4.66 \\
\#30  &  BAM:0036 &      H4 &   2.03 & 2.70 & 1.00 &       207 &                56.85 &             15.20 & 2526 &   5.40 \\
\#31  &  BAM:0046 &      H4 &   2.03 & 2.70 & 1.16 &       210 &                59.62 &             13.42 & 2344 &   3.52 \\
\rowcolor{gray!25}\#32  &  BAM:0048 &      H4 &   2.03 & 2.75 & 1.25 &       191 &                54.30 &             13.08 & 2416 &   3.46 \\
\#33  &  BAM:0053 &      H4 &   2.03 & 2.75 & 1.50 &       205 &                58.18 &              7.98 & 2471 &   3.14 \\
\rowcolor{gray!25}\#34  &  BAM:0057 &      H4 &   2.03 & 2.75 & 1.75 &       223 &               105.69 &              2.19 & 2490 &   0.90 \\
\#35  &  BAM:0058 &    MPA1 &   2.47 & 2.70 & 1.00 &       114 &                49.16 &             19.34 & 2720 &   2.23 \\
\#36  &  BAM:0059 &     MS1 &   2.77 & 2.70 & 1.16 &       328 &                67.56 &             15.08 & 2065 &   3.32 \\
\#37  &  BAM:0061 &     MS1 &   2.77 & 2.70 & 1.00 &       323 &                67.84 &             14.13 & 2013 &   2.47 \\
\#38  &  BAM:0065 &    MS1b &   2.76 & 2.70 & 1.00 &       287 &                62.25 &             18.77 & 2043 &   3.10 \\
\#39  &  BAM:0070 &    MS1b &   2.76 & 2.75 & 1.00 &       260 &                61.82 &             14.85 & 2120 &   3.52 \\
\#40  &  BAM:0080 &    MS1b &   2.76 & 2.50 & 1.27 &       439 &                76.67 &             11.83 & 2084 &   3.61 \\
\#41  &  BAM:0089 &    MS1b &   2.76 & 2.75 & 1.25 &       266 &                65.45 &             12.67 & 2067 &   2.97 \\
\#42  &  BAM:0090 &    MS1b &   2.76 & 3.20 & 1.00 &       112 &                48.33 &             17.35 & 2306 &   4.26 \\
\#43  &  BAM:0091 &    MS1b &   2.76 & 2.75 & 1.50 &       278 &                59.63 &              8.37 & 1956 &   2.14 \\
\#44  &  BAM:0092 &    MS1b &   2.76 & 3.40 & 1.00 &        79 &                41.57 &             19.48 & 2433 &   4.34 \\
\#45  &  BAM:0093 &    MS1b &   2.76 & 2.75 & 1.75 &       293 &                73.70 &              5.84 & 1970 &   1.55 \\
\#46  &  BAM:0098 &     SLy &   2.06 & 2.70 & 1.00 &        73 &                41.88 &             17.33 & 3340 &   3.29 \\
\#47  &  BAM:0107 &     SLy &   2.06 & 2.46 & 1.22 &       135 &                49.78 &             13.86 & 2784 &   3.03 \\
\#48  &  BAM:0121 &     SLy &   2.06 & 2.50 & 1.27 &       124 &                49.16 &             15.10 & 2787 &   3.89 \\
\#49  &  BAM:0122 &     SLy &   2.06 & 2.60 & 1.17 &        95 &                45.95 &             13.03 & 3050 &   3.87 \\
\#50  &  BAM:0123 &     SLy &   2.06 & 2.70 & 1.16 &        74 &                41.61 &             17.79 & 3362 &   3.31 \\
\#51  &  BAM:0124 &     SLy &   2.06 & 2.50 & 1.50 &       134 &                50.27 &              8.96 & 2951 &   2.95 \\
\#52  &  BAM:0126 &     SLy &   2.06 & 2.75 & 1.25 &        68 &                41.66 &             12.50 & 3460 &   1.89 \\
\#53  &  BAM:0128 &     SLy &   2.06 & 2.75 & 1.50 &        76 &                45.73 &              6.60 & 3339 &   1.45 \\
\#54  &    whisky &  Gamma2 &   1.82 & 2.90 & 1.00 &       277 &                   -  &                -  & 2127 &     -  \\
\#55  &    whisky &  Gamma2 &   1.82 & 2.85 & 1.00 &       324 &                   -  &                -  & 2183 &     -  \\
\#56  &    whisky &  Gamma2 &   1.82 & 2.80 & 1.00 &       379 &                   -  &                -  & 2061 &     -  \\
\#57  &    whisky &  Gamma2 &   1.82 & 2.75 & 1.00 &       442 &                   -  &                -  & 2004 &     -  \\
\#58  &    whisky &  Gamma2 &   1.82 & 2.70 & 1.00 &       516 &                   -  &                -  & 1930 &     -  \\
\#59  &    whisky &    GNH3 &   1.98 & 2.50 & 1.00 &       345 &                   -  &                -  & 2272 &     -  \\
\#60  &    whisky &    GNH3 &   1.98 & 2.55 & 1.00 &       306 &                   -  &                -  & 2302 &     -  \\
\#61  &    whisky &    GNH3 &   1.98 & 2.60 & 1.00 &       271 &                   -  &                -  & 2425 &     -  \\
\#62  &    whisky &    GNH3 &   1.98 & 2.65 & 1.00 &       240 &                   -  &                -  & 2479 &     -  \\
\#63  &    whisky &    GNH3 &   1.98 & 2.70 & 1.00 &       213 &                   -  &                -  & 2595 &     -  \\
\#64  &    whisky &    ALF2 &   1.99 & 2.45 & 1.00 &       236 &                   -  &                -  & 2443 &     -  \\
\#65  &    whisky &    ALF2 &   1.99 & 2.50 & 1.00 &       212 &                   -  &                -  & 2493 &     -  \\
\#66  &    whisky &    ALF2 &   1.99 & 2.55 & 1.00 &       190 &                   -  &                -  & 2574 &     -  \\
\#67  &    whisky &    ALF2 &   1.99 & 2.60 & 1.00 &       170 &                   -  &                -  & 2655 &     -  \\
\#68  &    whisky &    ALF2 &   1.99 & 2.65 & 1.00 &       153 &                   -  &                -  & 2693 &     -  \\
\#69  &    whisky &      H4 &   2.03 & 2.50 & 1.00 &       327 &                   -  &                -  & 2247 &     -  \\
\#70  &    whisky &      H4 &   2.03 & 2.55 & 1.00 &       292 &                   -  &                -  & 2377 &     -  \\
\#71  &    whisky &      H4 &   2.03 & 2.60 & 1.00 &       260 &                   -  &                -  & 2356 &     -  \\
\#72  &    whisky &      H4 &   2.03 & 2.65 & 1.00 &       232 &                   -  &                -  & 2449 &     -  \\
\#73  &    whisky &      H4 &   2.03 & 2.70 & 1.00 &       208 &                   -  &                -  & 2501 &     -  \\
\#74  &    whisky &     SLy &   2.06 & 2.50 & 1.00 &       118 &                   -  &                -  & 3154 &     -  \\
\#75  &    whisky &     SLy &   2.06 & 2.55 & 1.00 &       105 &                   -  &                -  & 3235 &     -  \\
\#76  &    whisky &     SLy &   2.06 & 2.60 & 1.00 &        93 &                   -  &                -  & 3229 &     -  \\
\#77  &    whisky &     SLy &   2.06 & 2.65 & 1.00 &        82 &                   -  &                -  & 3282 &     -  \\
\#78  &    whisky &     SLy &   2.06 & 2.70 & 1.00 &        73 &                   -  &                -  & 3338 &     -  \\
\#79  &    whisky &     SLy &   2.06 & 2.60 & 1.08 &        93 &                   -  &                -  & 3212 &     -  \\
\#80  &    whisky &    APR4 &   2.20 & 2.55 & 1.00 &        85 &                   -  &                -  & 3229 &     -  \\
\#81  &    whisky &    APR4 &   2.20 & 2.60 & 1.00 &        75 &                   -  &                -  & 3279 &     -  \\
\#82  &    whisky &    APR4 &   2.20 & 2.65 & 1.00 &        67 &                   -  &                -  & 3373 &     -  \\
\#83  &    whisky &    APR4 &   2.20 & 2.70 & 1.00 &        60 &                   -  &                -  & 3462 &     -  \\
\#84  &     sacra &    APR4 &   2.20 & 2.70 & 1.00 &        60 &                   -  &                -  & 3450 &     -  \\
\#85  &     sacra &    APR4 &   2.20 & 2.70 & 1.00 &        60 &                   -  &                -  & 3255 &     -  \\
\#86  &     sacra &    APR4 &   2.20 & 2.70 & 1.00 &        60 &                   -  &                -  & 3330 &     -  \\
\#87  &     sacra &    APR4 &   2.20 & 2.60 & 1.00 &        76 &                   -  &                -  & 3210 &     -  \\
\#88  &     sacra &     SLy &   2.06 & 2.70 & 1.25 &        76 &                   -  &                -  & 3340 &     -  \\
\#89  &     sacra &     SLy &   2.06 & 2.70 & 1.16 &        74 &                   -  &                -  & 3320 &     -  \\
\#90  &     sacra &     SLy &   2.06 & 2.70 & 1.08 &        73 &                   -  &                -  & 3390 &     -  \\
\#91  &     sacra &     SLy &   2.06 & 2.70 & 1.00 &        73 &                   -  &                -  & 3480 &     -  \\
\#92  &     sacra &     SLy &   2.06 & 2.60 & 1.00 &        93 &                   -  &                -  & 3160 &     -  \\
\#93  &     sacra &    ALF2 &   1.99 & 2.80 & 1.00 &       110 &                   -  &                -  & 2920 &     -  \\
\#94  &     sacra &    ALF2 &   1.99 & 2.70 & 1.25 &       139 &                   -  &                -  & 2820 &     -  \\
\#95  &     sacra &    ALF2 &   1.99 & 2.70 & 1.16 &       138 &                   -  &                -  & 2650 &     -  \\
\#96  &     sacra &    ALF2 &   1.99 & 2.70 & 1.08 &       137 &                   -  &                -  & 2770 &     -  \\
\#97  &     sacra &    ALF2 &   1.99 & 2.70 & 1.00 &       137 &                   -  &                -  & 2770 &     -  \\
\#98  &     sacra &    ALF2 &   1.99 & 2.60 & 1.00 &       170 &                   -  &                -  & 2630 &     -  \\
\#99  &     sacra &      H4 &   2.03 & 2.90 & 1.00 &       133 &                   -  &                -  & 2930 &     -  \\
\#100 &     sacra &      H4 &   2.03 & 2.80 & 1.15 &       168 &                   -  &                -  & 2505 &     -  \\
\#101 &     sacra &      H4 &   2.03 & 2.80 & 1.00 &       166 &                   -  &                -  & 2780 &     -  \\
\#102 &     sacra &      H4 &   2.03 & 2.70 & 1.25 &       214 &                   -  &                -  & 2320 &     -  \\
\#103 &     sacra &      H4 &   2.03 & 2.70 & 1.25 &       214 &                   -  &                -  & 2340 &     -  \\
\#104 &     sacra &      H4 &   2.03 & 2.70 & 1.25 &       214 &                   -  &                -  & 2300 &     -  \\
\#105 &     sacra &      H4 &   2.03 & 2.70 & 1.16 &       210 &                   -  &                -  & 2440 &     -  \\
\#106 &     sacra &      H4 &   2.03 & 2.70 & 1.08 &       208 &                   -  &                -  & 2475 &     -  \\
\#107 &     sacra &      H4 &   2.03 & 2.70 & 1.00 &       208 &                   -  &                -  & 2590 &     -  \\
\#108 &     sacra &      H4 &   2.03 & 2.70 & 1.00 &       208 &                   -  &                -  & 2530 &     -  \\
\#109 &     sacra &      H4 &   2.03 & 2.70 & 1.00 &       208 &                   -  &                -  & 2490 &     -  \\
\#110 &     sacra &      H4 &   2.03 & 2.60 & 1.17 &       264 &                   -  &                -  & 2370 &     -  \\
\#111 &     sacra &      H4 &   2.03 & 2.60 & 1.08 &       261 &                   -  &                -  & 2260 &     -  \\
\#112 &     sacra &      H4 &   2.03 & 2.60 & 1.00 &       260 &                   -  &                -  & 2310 &     -  \\
\#113 &     sacra &     MS1 &   2.77 & 2.90 & 1.23 &       224 &                   -  &                -  & 2120 &     -  \\
\#114 &     sacra &     MS1 &   2.77 & 2.90 & 1.00 &       219 &                   -  &                -  & 2110 &     -  \\
\#115 &     sacra &     MS1 &   2.77 & 2.80 & 1.00 &       266 &                   -  &                -  & 2045 &     -  \\
\#116 &     sacra &     MS1 &   2.77 & 2.70 & 1.25 &       332 &                   -  &                -  & 2110 &     -  \\
\#117 &     sacra &     MS1 &   2.77 & 2.70 & 1.16 &       328 &                   -  &                -  & 2050 &     -  \\
\#118 &     sacra &     MS1 &   2.77 & 2.70 & 1.08 &       326 &                   -  &                -  & 2050 &     -  \\
\#119 &     sacra &     MS1 &   2.77 & 2.70 & 1.00 &       325 &                   -  &                -  & 2020 &     -  \\
\#120 &     sacra &     MS1 &   2.77 & 2.60 & 1.00 &       398 &                   -  &                -  & 1960 &     -  \\
\end{longtable}
\end{center}

\end{document}